\def\draftversion{false}
\RequirePackage{ifthen}
\ifthenelse{\equal{\draftversion}{true}}{
\documentclass[floatfix,galley,prb]{revtex4}
}{
\documentclass[floatfix, twocolumn, prb]{revtex4}
}
\usepackage{hyperref}
\usepackage{color}
\usepackage{multirow}
\usepackage{epsfig}
\usepackage{graphicx}
\usepackage{amsmath}
\usepackage{amssymb}
\usepackage{bm}
\usepackage{dcolumn}
\usepackage{braket}
\usepackage{bbold}
\usepackage{ulem}

\begin{document}

\title{Metal-insulator transition and topological properties of pyrochlore iridates}

\author{Hongbin~Zhang}
\author{Kristjan~Haule}
\author{David~Vanderbilt}
\affiliation{Department of Physics and Astronomy, Rutgers
University, Piscataway, USA}

\date{April 8, 2015}

\begin{abstract}
Combining density functional theory
(DFT) and embedded dynamical mean-field theory (DMFT)
methods, we study the metal-insulator transition in
$R_2$Ir$_2$O$_7$ ($R$=Y, Eu, Sm, Nd, Pr, and Bi) and the topological
nature of the insulating compounds.  Accurate free energies
evaluated using the charge self-consistent DFT+DMFT method reveal that
the metal-insulator transition occurs for an A-cation radius between
that of Nd and Pr, in agreement with experiments.
The all-in-all-out
magnetic phase, which is stable in the Nd compound but not the Pr
one, gives rise to a small Ir$^{4+}$ magnetic moment of
$\approx 0.5\mu_B$ and opens a sizable correlated gap.
We demonstrate that within this state-of-the-art theoretical method,
the insulating bulk pyrochlore iridates are topologically trivial.
\end{abstract}

\maketitle

\def\scr{\scriptsize}
\ifthenelse{\equal{\draftversion}{true}}{
\marginparwidth 2.7in
\marginparsep 0.5in
\newcounter{comm} 
\def\commnext{\stepcounter{comm}}
\def\commtext{{\bf\color{blue}[\arabic{comm}]}}
\def\commmar{{\bf\color{blue}[\arabic{comm}]}}
\def\dvm#1{\commnext\marginpar{\small DV\commmar: #1}\commtext}
\def\hzm#1{\commnext\marginpar{\small HZ\commmar: #1}\commtext}
\def\khm#1{\commnext\marginpar{\small KH\commmar: #1}\commtext}
}{
\def\dvm#1{}
\def\hzm#1{}
\def\khm#1{}
}

Interest in $5d$ compounds has blossomed in recent years, as they
provide a promising playground for interesting new
physics\cite{Kim:2008, Kim:2009, Jackeli:2009, Shitade:2009, Wang:2011}
arising from the interplay of the atomic spin-orbit coupling (SOC),
which scales as $Z^4$ with the atomic number $Z$, and the electronic
correlations, which are reduced due to the spatially more
extended wavefunctions of the $5d$ ions.
Rare-earth pyrochlore iridates ($R_2$Ir$_2$O$_7$)
have drawn intensive
attention in recent years~\cite{Krempa:2014} because of their
geometrically frustrated lattice, which favors the spin-liquid
phase,~\cite{Tokiwa:2014} and the possibility to host various
nontrivial topological phases in the bulk,\cite{Pesin:2010, Yang:2010,
Kargarian:2011, Wan:2011, Go:2012} thin films,\cite{Hu:2012,
Yang:2014} and domain walls.\cite{Yamaji:2014}
Depending on the radius of the A-site cation, rare-earth pyrochlore
iridates undergo a metal-insulator transition (MIT),~\cite{Matsuhira:2011}
concomitant with a transition to an all-in-all-out (AIAO) magnetic
state.~\cite{Disseler:2012a, Shapiro:2012, Tomiyasu:2012}
Nevertheless, when the A-site cation is Pr or the more covalent Bi, the
corresponding pyrocholore iridates are metallic down to very low
temperature with no long-range magnetic ordering.~\cite{Tokiwa:2014,
Qi:2012}

Recently, inspired by pioneering work based on a generic tight-binding
model,~\cite{Pesin:2010} there have been many studies focusing on
nontrivial topological phases in bulk pyrochlore
iridates.~\cite{Yang:2010, Kargarian:2011, Wan:2011, Go:2012}
Based on calculations using the local density approximation
(LDA) with Hubbard $U$ (LDA+U) including SOC,
Wan~{\it et al.}~predicted that Y$_2$Ir$_2$O$_7$ can
host nontrivial Weyl semimetal and axion insulator
phases.~\cite{Wan:2011}
Using a simplified single-band Hubbard model on the pyrochlore lattice,
Go~{\it et al.}~conducted
cluster-DMFT calculations~\cite{Go:2012} and confirmed the LDA+U+SOC phase
diagram, but with a magnetic configuration having the Ir magnetic moments
rotated by 90$^\circ$ from the AIAO ground-state configuration.

Iridates are known to be beyond the range of applicability
of LDA+U calculations, since dynamical local correlations are
crucial to explain their complicated electronic structure.
Instead, the LDA+DMFT method~\cite{Zhang:2013} correctly gives the
boundary of the MIT in the Ruddlesden-Popper (RP) series,
and the fine details of the
ARPES measurements of Sr$_2$IrO$_4$~\cite{Wang:2013} can only be
understood by comparing with the
LDA+DMFT spectral functions, which inlcude essentially exact
correlation effects local to a given Ir atom.~\cite{Zhang:2013}
Moreover, the LDA+U method overestimates the stability of the
insulating magnetically ordered states, and the value of $U$, which
is to some extent an adjustable parameter, often needs to be
adjusted for each member of the series. Thus, a consistent
treatment of the electronic correlations together with spin-orbit
coupling at the LDA+DMFT level is necessary to shed light on the MIT
in the pyrochlore iridates.

In this work, we have carried out all-electron charge self-consistent
LDA+DMFT~\cite{Haule:2010} calculations. We have studied a series of
pyrochlore iridates in both the paramagnetic (PM) and AIAO magnetic states,
focusing on the metal-insulator transition with respect to the
A-cation radius and the topological nature of the insulating
states.
We demonstrate that the MIT occurs in those
compounds which can host an AIAO magnetic ground state.
Because of the large degree of geometric frustration in the
pyrochlore lattice, a significant quasiparticle mass
in the PM state is needed to destabilize the Fermi liquid formation
at the expense of magnetic long-range order.  The tuning across the
boundary between the Kondo-screened Fermi-liquid solution and the
magnetic AIAO solution is achived with A-cation substitution. The most
important consequence of the latter is the change of oxygen
coordinates, which results in a slightly different Ir hybridization and
effective Ir-$t_{2g}$ bandwidth.
Based on calculations of the wavefunction parities in a many-body context,
we conclude that the insulating pyrochlore iridates are likely
to remain topologically trivial.

Our LDA+DMFT
calculations were done using the projection/embedding
implementation,\cite{Haule:2010} which avoids downfolding or model
building, and adds dynamic local correlations to a set of localized
quasi-atomic orbitals. The DFT part is implemented using the WIEN2k
package.~\cite{wien2k}
All parameters of the calculation are identical to
those used in our earlier work on RP
iridates.\cite{Zhang:2013}
We also performed LDA+U calculations using the VASP
code~\cite{vasp} for comparison.  Details are provided in the
Supplemental Materials.

\begin{figure}[!h]
\centering
\includegraphics[width=7.0cm]{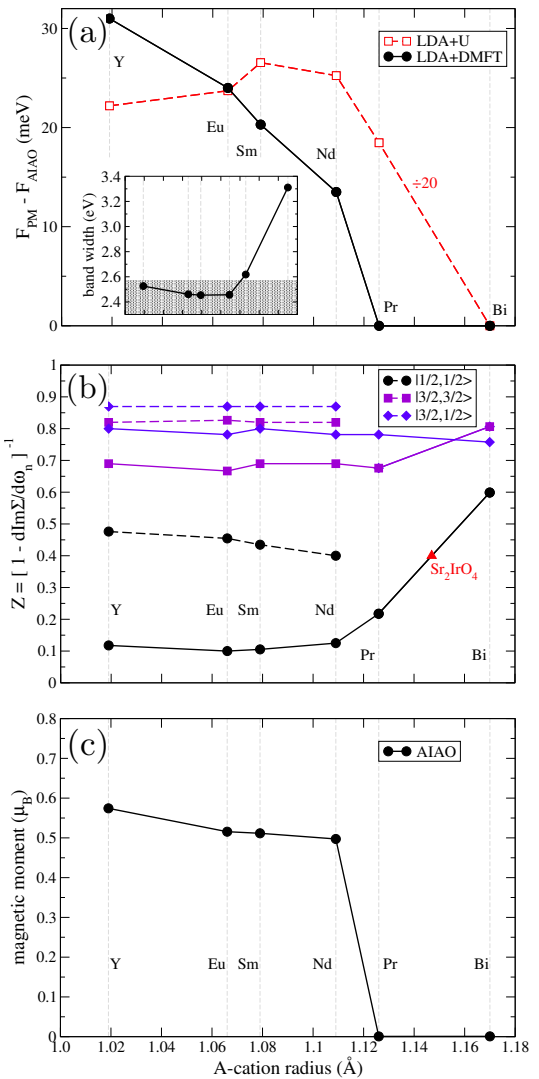}
\caption{(Color online) Variation with respect to the A-cation radius of
computed properties of $R_2$Ir$_2$O$_7$ ($R$=Y, Eu, Sm, Nd, Pr, and Bi).
(a) Difference of free energies $F$ between PM and AIAO magnetic states
for $R_2$Ir$_2$O$_7$.  Solid circles
(hollow squares) denote the values obtained using the LDA+DMFT (LDA+U)
method, plotted as zero when the AIAO state cannot be stabilized.
Inset displays the LDA band-width of the $t_{2g}$ bands;
shaded region highlights the insulating compounds.
(b) Quasiparticle renormalization amplitude $Z_\alpha$
(see text for the definition) for the $J_\text{eff}\!=\!1/2$ and
$J_\text{eff}\!=\!3/2$ states.  Solid (dashed) lines
denote the values in the PM (AIAO) states, with $Z$ for the
$J_\text{eff}\!=\!1/2$ orbital in the PM state of Sr$_2$IrO$_4$
(red triangle) shown for comparison.
(c) Ordered moment of Ir atoms.
All LDA+DMFT calculations are done at $50$K.}
\label{fig:mit}
\end{figure}

For the series of $R_2$Ir$_2$O$_7$ ($R$=Y, Eu, Sm, Nd, Pr, and Bi)
compounds, we considered the PM and AIAO magnetic solutions.
We found that the magnetic AIAO solution could be stabilized
(that is, we could find a stable or metastable solution) only in
the Y, Eu, Sm, and Nd compounds, and was always insulating (see
spectral functions in the Supplement), while the PM solution could be
stabilized for all materials and was always metallic, consistent with
experiments.~\cite{Matsuhira:2011, Qi:2012}
Naively, the MIT in the six $R_2$Ir$_2$O$_7$ compounds can be understood
in terms of the
bandwidth of the Ir-$t_{2g}$ bands, as shown as inset of Fig.~\ref{fig:mit}(a).
For instance, at the LDA level
there is a critical value of about $2.6$~eV for the Ir-$t_{2g}$ bandwidth
separating
the compounds with insulating and metallic ground states.
A detailed analysis reveals that the bandwidth is closely correlated
with the hybridization function, and its variation can be attributed
to the change of the oxygen coordinates (see the Supplement).

To determine the relative stability of the two phases, we computed the
free energy differences between PM and AIAO states following
Ref.~\onlinecite{Haule:2015}, yielding the results
displayed in Fig.~\ref{fig:mit}(a).
For comparison, the difference of LDA+U total energies is also
shown, but reduced by a factor of 20 to fit in the plot.
For each method, we used the same $U$ as determined in our previous study
of RP iridates.~\cite{Zhang:2013}
The LDA+U energy difference is about $400$\,meV on average,
much too large to be compatible with the known ordering
temperatures. This is hardly surprising since the local moments are
zero in the PM state within LDA+U, which cannot describe fluctuating
local moments. As a result, the energy of the PM state is
severely overestimated.

Experimentally, Pr$_2$Ir$_2$O$_7$ has a non-magnetic metallic ground
state,\cite{Matsuhira:2011} correctly described by LDA+DMFT.
In LDA+U by contrast, only Bi$_2$Ir$_2$O$_7$ is metallic, while the ground
state of Pr$_2$Ir$_2$O$_7$ is an AIAO magnetic insulator.
We also studied other possible magnetic phases within LDA+DMFT. We
found that the three-in-one-out configuration can also be stabilized
for some compounds, but is higher in energy than the AIAO state
({\it e.g.}, by $23$~meV for Y$_2$Ir$_2$O$_7$). The two-in-two-out state, on
the other hand, is not stable in this method.  Therefore, the AIAO
magnetic state is indeed the ground state of insulating pyrochlore
iridates within LDA+DMFT.

Our self-consistent LDA+DMFT calculations also give reasonable
estimates of the free-energy differences between PM and AIAO states
as shown in Fig.~\ref{fig:mit}(a).  The average free-energy difference is about
$20$\,meV, which is about twice the typical magnetic ordering
temperature (about $100$\,K) observed
experimentally.~\cite{Matsuhira:2011}
Moreover, the trend of the LDA+DMFT free-energy difference with
respect to the A-cation radius follows loosely the experimental
magnetic ordering temperature.~\cite{Matsuhira:2011, Qi:2012}
We note in passing that the difference in electronic entropy $S$ between the
two phases is very small. For example, in Nd$_2$Ir$_2$O$_7$ we obtained
$\Delta S\!=\!0.033\,k_B$ per Ir atom, which is similar to the
experimentally determined $\Delta S\!=\!0.028\,k_B$~\cite{Matsuhira:2011}
(see the Supplement).

To shed more light on the origin of the MIT,
Fig.~\ref{fig:mit}(b) shows the DMFT-predicted quasiparticle
renormalization amplitude $Z_{\alpha}\!=\!\left(1-d\,\text{Im}
{\Sigma_{\alpha\alpha}}/{d\omega_n}\right)^{-1}|_{\omega_n=0}$,
where $\alpha$ labels a correlated orbital and $\Sigma$ indicates
the self-energy on the Matsubara-frequency ($\omega_n$) axis.
The less active $J_\text{eff}\!=\!3/2$ states are
only weakly renormalized. 
Turning to the more active $J_\text{eff}\!=\!1/2$ states,
these are renormalized by a factor
of almost ten in the PM state for compounds having an insulating
ground state. By contrast, the $J_\text{eff}\!=\!1/2$
quasiparticles in Bi$_2$Ir$_2$O$_7$ remain quite
light, consistent with recent ARPES measurement.~\cite{Wang:2015}
Surprisingly, the $J_\text{eff}\!=\!1/2$ orbital in
Pr$_2$Ir$_2$O$_7$ is
quite heavy (about five times the band mass), indicating that the
correlations are considerably stronger
than in the marginal Mott insulator Sr$_2$IrO$_4$,
whose $Z_{\alpha}$ is plotted as
the red triangle in Fig.~\ref{fig:mit}(b).
In this sense, the frustration in the pyrochlore lattice plays an
essential role in shifting the MIT boundary towards stronger
correlations compared to unfrustrated RP iridates. The frustration
thus promotes metallicity by penalizing competing long-range
magnetic order, an
effect which is well captured by DMFT but less adequately by LDA+U,
explaining why LDA+U predicts the Pr compound to be insulating.

In short,
the boundary of the MIT across the $R_2$Ir$_2$O$_7$
compounds is determined by a competition between the formation of
a quasiparticle band via the Kondo effect in the PM state, and the
tendency toward long-range magnetic order.
Both can lead to an effective reduction of the free energy, and are favored
by enhanced quantum fluctuations.
In general, the occurrence of a MIT depends crucially on many factors,
including dimensionality and frustration.
For example, in Sr$_2$IrO$_4$ the two-dimensionality
reduces the N\'eel ordering temperature, but the short-range
order above the N\'eel temperature still preserves the charge gap in
the excitation spectrum. Frustration also reduces the tendency to
long-range order, but it promotes metallicity and allows very narrow
quasiparticles to be observed in the PM state before the long-range
order opens the charge gap. This is why Pr$_2$Ir$_2$O$_7$
remains in the paramagnetic bad-metal phase even though the
correlation strength is much larger than in Sr$_2$IrO$_4$.

For correlated Fermi liquids, the quantum fluctuations give rise to
fluctuating moments in the PM state, which statically order when
the magnetic energy gain is sufficient to overcome the gain available
from band formation.
According to Fig.~\ref{fig:mit}(c), the magnitude of the ordered
magnetic moment for Y$_2$Ir$_2$O$_7$ is
about $0.57\,\mu_B$.  This is a bit larger than the
experimental upper bound of $0.5\,\mu_B$ obtained by neutron
scattering,~\cite{Shapiro:2012} which can be attributed to the fact
that the spatial fluctuations, which are not fully accounted for in
LDA+DMFT, are likely to further reduce the ordered magnetic moment.
In the AIAO insulating state, there are no states at the Fermi energy,
but the slope of the self-energy at zero frequency still gives some
measure of the importance of quantum fluctuations. As shown by the dashed
lines in Fig.~\ref{fig:mit}(b), the quantum renormalization effects
are greatly reduced in the AIAO state as compared to the PM state.

We observed that,
like the effective $J_\text{eff}\!=\!1/2$ states in the RP iridates,~\cite{Zhang:2013}
the $J_\text{eff}\!=\!1/2$ wavefunction in the pyrochlore iridates also
strongly deviates from the rotationally invariant SU(2) point. This is
mostly due to the trigonal crystal field induced by compression of
the IrO$_6$ octahedra.  The trigonal crystal-field splitting in our
calculations is derived from a very large energy window,
and is on average about $0.39$\,eV (see the Supplement),
in good agreement with resonant X-ray scattering
measurements.~\cite{Hozoi:2012}
Correspondingly, the resulting orbital-to-spin moment ratio is about
1.3 (see the Supplement), showing a significant deviation from the $SU(2)$
limit of two.~\cite{Zhang:2013}
Nevertheless, the effective $J_\text{eff}\!=\!1/2$ and
$J_\text{eff}\!=\!3/2$ bands are
still well separated in energy, which facilitates our analysis of the
topological character. 
Note that the values of local $U$ and $J$ parameters 
on Ir atoms in our LDA+DMFT calculations are the same for the pyrochlore
and RP iridates,~\cite{Zhang:2013}
confirming the more universal nature of
local Coulomb repulsion across similar materials when the
screening by hybridization effects at high energy are allowed in the
calculation.

\begin{figure}[!h]
\centering
\includegraphics[width=7.0cm]{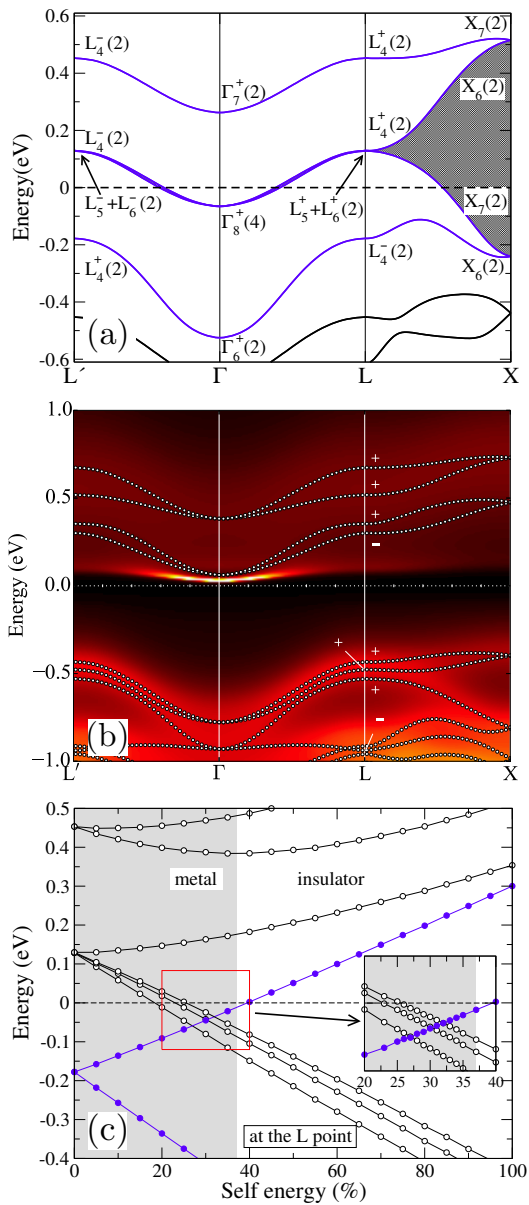}
\caption{(Color online)
(a) The LDA-like band structure of Nd$_2$Ir$_2$O$_7$ with irreducible
representations indicated at the time-reversal invariant momenta;
numbers in parenthesis denote degeneracies.
(b) Color map: spectral function of Nd$_2$Ir$_2$O$_7$ in the AIAO
magnetic state. Dotted lines: band structure of the
effective topological Hamiltonian (see text for details).
The parities of eight $J_\text{eff}\!=\!1/2$-derived
bands are indicated by ``$+$" or ``$-$" at the $L$ point.
(c) Evolution of the eigenvalues of the
effective topological Hamiltonian at $L$ with
mixing fraction of the DMFT self-energy.
Inset zooms in on the critical region indicated by the red square.
}\label{fig:topo}
\end{figure}

We now turn to the topological properties of the insulating
pyrochlores.  Fig.~\ref{fig:topo}(a) shows a fictitious non-magnetic
band structure of Nd$_2$Ir$_2$O$_7$, together with the irreducible
representations for the $J_\text{eff}\!=\!1/2$ states around $E_F$ at the
eight time-reversal-invariant momenta (TRIM). This band structure is
obtained by neglecting the dynamic part of the LDA+DMFT self-energy,
but using the self-consistent LDA+DMFT charge density, with
time-reversal (TR) symmetry imposed. We note that these bands are
quite similar to the LDA bands.

Regarding the irreducible representations, the space group of the
pyrochlore iridates is nonsymmorphic.  Thus, depending on the choice
of inversion center, the four energetically equivalent $L$ points are
separated into one $L^\prime$ and three $L$ points, with opposite
parities for the $L^\prime$ and $L$ points as shown in
Fig.~\ref{fig:topo}(a).
Moreover, at each of the three X points all states are four-fold
degenerate by symmetry,
with each degenerate group comprising two even-parity and two
odd-party states.
Finally, all eight $J_\text{eff}\!=\!1/2$ bands at the $\Gamma$ point
have even parity.
The situation is summarized in the PM column of Table~\ref{table:parity}.
It follows that if a global band gap could be opened so as to extend
the shaded region of Fig.~\ref{fig:topo}(a) throughout the Brillouin
zone, corresponding to a half-filled $J_\text{eff}\!=\!1/2$ manifold,
seven out of 16 of the occupied Kramers pairs at the eight TRIM would
be odd-parity ones. Since this number is odd, it would generate a
strong-topological-insulator phase.~\cite{Fu:2007} It has
been suggested this might be achieved for Pr$_2$Ir$_2$O$_7$, for
example, by applying a strain along the (111) direction.~\cite{Chen:2015}
If the TR symmetry would be weakly broken in this phase, the system
would become an axion rather than a strong topological insulator, but
we shall use the term ``topological insulator'' to cover both cases.

%
\begin{table}
\caption{Parity analysis of eight eigenstates around E$_F$ in the PM
and AIAO magnetic states, corresponding to Figs.~\ref{fig:topo}(a-b)
respectively.  `MULT' denotes the multiplicity in accounting for the
eight time-reversal invariant momenta.
The parities are given in order of increasing energy
eigenvalues, with vertical bars indicating $E_F$.}
\begin{tabular}{c|c|c|c}
\hline\hline
&    MULT           &    PM             &     AIAO  \\  \hline
$\Gamma$       &         1               & \{$+ + + + \mid + + +  + $\}  &  \{$ + + + + \mid + + + + $\}  \\
$L$                    &         3               & \{$ - - + + \mid + + + + $\}    & \{$ - + + + \mid - + + + $\}  \\
$L^\prime $      &         1               & \{$ + + - - \mid - - - -$\}       &   \{$ + - - -\mid + - - -$\} \\
$X$                    &         3               &  \{$ + + - -\mid + + - -$\}     &   \{$ + + - -\mid + + - -$\} \\ \hline
\end{tabular}\label{table:parity}
\end{table}

When the DMFT self-energy is taken into account, the TR
symmetry is spontaneously broken and a global gap opens up.
However, the topological character may be different from
that described above, because of the drastic mixing of bands that results.
Following the theory of interacting topological
phases,~\cite{Wang:2012,Turner:2012} the topological indices
can be obtained by inspecting an effective single-particle
Hamiltonian defined as $H_\text{eff}\!=\!H_0+\Sigma(\omega\!=\!0)$,
where $H_0$ is the TR-invariant Bloch Hamiltonian and
$\Sigma(\omega\!=\!0)$ is the DMFT self-energy at zero frequency
(i.e., at the Fermi energy), which carries all the TR-breaking
terms.
The resulting band structure is shown in Fig.~\ref{fig:topo}(b),
together with the full LDA+DMFT spectral function.  Since TR
symmetry is broken, the total parity of all occupied eigenstates
at the TRIM should now be counted;\cite{Turner:2012} the system will
be a trivial or axion insulator if the total number is $4n$ or
$4n+2$ respectively, with $n$ an integer.

The parities of the $J_\text{eff}\!=\!1/2$-derived eigenstates
at the eight TRIM are also shown for the AIAO state
in Table~\ref{table:parity}.  At the
$\Gamma$ point all states remain even-parity even with the
addition of the self-energy.
At the $X$ points, each four-fold degenerate state splits into two
doublets, and we find (see the Supplement) that including $\Sigma(\omega\!=\!0)$
does not induce any band inversion or otherwise change the
ordering of parities, so that
the number of occupied odd-parity states remains equal to two at
each $X$ point.
The most significant change occurs at the $L$ ($L^\prime$) points,
where the TR-symmetry breaking splits each doubly-degenerate
$L_4^-$ ($L_4^+$) state into two singlets. Some of these cross the gap
as the self-energy is turned on, as shown in
Fig.~\ref{fig:topo}(c), with the result that there is an exchange
of parity between occupied and unoccupied states
as shown in Table~\ref{table:parity}.
As a result, there are a total of twelve odd-parity occupied states at
the TRIM. 
Similar behavior is also observed in the other insulating $R_2$Ir$_2$O$_7$
compounds studied.
Thus, after properly including the DMFT self-energy, we find that
the insulating $R_2$Ir$_2$O$_7$
compounds are topologically trivial insulators.  This is in agreement
with previous LDA+U results~\cite{Wan:2011} in the large-$U$ limit.

Nevertheless, in Refs.~\onlinecite{Wan:2011} and \onlinecite{Go:2012}
it was argued that nontrivial axion insulator and Weyl
semimetal phases occur with smaller values of on-site $U$.  This is not
supported by our DMFT calculations.  For instance, for
Y$_2$Ir$_2$O$_7$, which is the most insulating of the six compounds, the PM
solution is more stable than the AIAO magnetic state when the on-site
$U$ value is reduced to $4.0$\,eV (see the Supplement).
Moreover, as shown in Fig.~\ref{fig:topo}(c), we observe that the
parity exchange between occupied and unoccupied states at the $L$
and $L^\prime$ points happens before a global band gap is opened.
Therefore, we think it is very unlikely that a topological insulating
phase can be found in the bulk pyrochlore iridates.

In summary, our calculations provide a clear picture of the origin of
the MIT in pyrochlore iridates and its variation with respect to the A-cation
radius.  Moreover, our parity analysis of the many-body effective
$J_\text{eff}\!=\!1/2$ wave functions reveals that when insulating
magnetic phases appear in these pyrochlore iridates, they are very
likely to remain topologically trivial.

\acknowledgments
We acknowledge Gregory A. Fiete, Victor Chua, Ru Chen, S.M. Disseler, Xiang Hu, Zhicheng Zhong,
Jinjian Zhou, Yugui Yao, Jianpeng Liu, and Klaus Koepernik for helpful discussions.
This work was supported by NSF Grant DMREF-12-33349.

Note: As we were finalizing this manuscript a preprint
appeared~\cite{Hiroshi:2015} which presents
results for Y$_2$Ir$_2$O$_7$ that are similar to our own in some
respects. However, we do not agree that the magnetic moments on the
Ir atoms in Y$_2$Ir$_2$O$_7$ can be larger than $1.0\,\mu_B$
(see the Supplement for a detailed discussion).

\end{document}